\documentclass[preprint,showpacs,preprintnumbers,amsmath,amssymb]{revtex4}

\usepackage{graphicx}
\usepackage{dcolumn}
\usepackage{bm}

\begin{document}

\title{Shear viscosity and Bose statistics: \\ Capillary flow above $\lambda$ point}

\author{Shun-ichiro Koh}
 
 \email{koh@kochi-u.ac.jp}
\affiliation{ Physics Division, Faculty of Education, Kochi University  \\
        Akebono-cho, 2-5-1, Kochi, 780, Japan 
}%

\date{\today}

\begin{abstract}
The  gradual fall of the shear viscosity at $T_{\lambda}<T<2.8K$, 
observed in a liquid helium 4 flowing through a capillary,  is examined. 
 The disappearance of the shear viscosity in a capillary flow is a 
 manifestation of superfluidity in dissipative phenomena, the onset 
 mechanism of which is a subtle problem compared to that of  
 superfluidity in non-dissipative phenomena.  Applying the 
 linear-response theory to the reciprocal of the shear viscosity 
 coefficient $1/\eta $,  we relate these two types 
 of superfluidity  using the Kramers-Kronig relation.
 As a result, we obtain a formula describing the influence of 
 Bose statistics on the kinematic shear viscosity in terms of the 
 susceptibility, without getting involved in the details of 
 the dissipation mechanism of a liquid.  Compared to an ordinary liquid, 
a liquid helium 4 above $T_{\lambda}$ has a  $10^{-3}$ times smaller   
shear viscosity coefficient. Hence, although in the normal phase, 
it is already an anomalous liquid  under the 
strong influence of Bose statistics. The coherent many-body wave function grows to an 
 intermediate size between a macroscopic and a microscopic one,  
 not as a thermal fluctuation but as a thermal equilibrium state. Beginning with  bosons  
without the condensate, we make a perturbation calculation  of its 
susceptibility with respect to the repulsive interaction.  Using the 
above formula on the kinematic shear viscosity, we examine how, with 
decreasing temperature, the growth of the coherent wave function 
 gradually suppresses the shear viscosity, and 
finally  leads to a frictionless flow at the $\lambda$-point. 
\end{abstract}

\maketitle

\section{\label{sec:level1}Introduction}

Superfluidity  of a liquid helium 4 was first discovered in a flow through a narrow channel and a  
capillary \cite {kap}. In an ordinary liquid, the shear viscosity causes 
 the shear stress $F_{xy}$ between two adjacent layers moving at 
different velocities 
\begin{equation}
 F_{xy}=\eta \frac{\partial v_x}{\partial y¥}¥,
	\label{¥}
\end{equation}¥
where $\eta$ is the shear viscosity coefficient. In a narrow pipe with an inside radius 
$a$ and a length $L$, the field of velocity driven by the pressure difference $\Delta P$ 
 has a form such as
\begin{equation}
 v_z(r)=\frac{a^2-r^2}{4\eta¥}¥\frac{\Delta P}{L¥}¥,
	\label{¥}
\end{equation}¥
where $r$ is a radius in the cylindrical coordinates  (Poiseuille flow).
In the superfluid phase of a liquid helium 4, even when the pressure difference vanishes ($\Delta P=0$), 
one observes a non-vanishing flow ($v_z(r) \ne 0$), hence $\eta =0$ in Eq.(2).

The shear viscosity of a liquid helium 4 has been 
subjected to considerable experimental and theoretical studies.
At $0<T<T_{\lambda}$, an oscillating or rotating  probe immersed in 
 a liquid helium 4 experiences the shear stress in a complicated manner
(viscosity paradox). At temperatures well below $T_{\lambda}$, various 
excitations of a liquid are strictly suppressed except for phonons and 
rotons.  Hence, it is natural to assume that phonons and rotons are  
responsible for the shear viscosity. By regarding these elementary excitations as a weakly 
interacting dilute Bose gas, the formula in the kinetic theory of gases  
well describes the shear viscosity  of a liquid at $T \ll T_{\lambda}$,  which is a   
 success of the quasi-particle picture  in the many-body theory \cite {kha}.

 \begin{figure}
\includegraphics [scale=0.5]{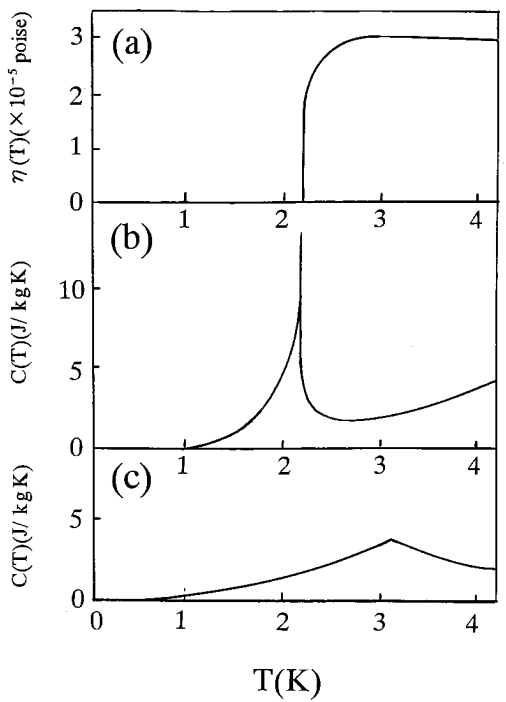}
\caption{\label{fig:epsart}(a) The temperature dependence of the shear viscosity 
$\eta (T)$ of a liquid helium 4 measured in 
 a flow through a capillary (after the data in Ref.3 and 5).
  (b) The temperature dependence of the specific 
 heat $C(T)$  of a liquid helium 4 (after Atkins, \protect{\it Liquid helium\/}4,
  (Cambridge,1959)).
  (c) $C(T)$  of an ideal Bose gas with the same density as a liquid helium 4.}
\end{figure}

For the shear viscosity at $T>T_{\lambda}$, however, the 
dilute-gas picture must not be applied, because the macroscopic 
condensate has not yet developed. Instead,  one must tackle the 
problem of how Bose statistics affects excitations and dissipations of a 
liquid. Figure 1(a) shows $\eta $ of a liquid helium 4 measured in a flow through a capillary 
at the vicinity of  $T_{\lambda}$ \cite {men}\cite {tje}\cite {zin}. One notes that  $\eta $ does not 
abruptly drops to zero at  $T_{\lambda}$, but it gradually decreases with decreasing 
temperature from $T_{\lambda}+0.7K$, in which the macroscopic condensate has not yet 
developed, and finally drops at the $\lambda$-point.

For the shear viscosity of a liquid, Maxwell obtained a simple formula 
$\eta=G\tau$ (Maxwell's relation), where $G$ is the 
modulus of rigidity and $\tau$ is a relaxation time,  using a physical 
argument \cite{max} (see Appendix.A). In the short time scale, a liquid 
shows similar behaviors to that of a solid at the molecular level.  
In a liquid, $G$ is determined  mainly by the motion of vacancy as well as in a 
solid. Since no apparent structural transformation is 
observed in a liquid helium 4 at the vicinity of $T_{\lambda}$,   $G$  
may be a constant at the  first approximation.  The gradual decrease of 
$\eta =G\tau $ above $T_{\lambda}$ suggests  that the influence of Bose 
statistics gives rise to a considerable decrease of $\tau$. 
To understand superfluidity occurring in such a 
dissipative flow, we must deal with the microscopic mechanism of the 
dissipations determining $\tau $. At first sight, it seems to be a hopeless attempt, because the dissipation 
itself is a complicated phenomenon  allowing no simple description \cite {han}. In a superfluid 
helium 4, many features associated to Bose statistics are masked by the 
strongly interacting character of the liquid. But, from a different viewpoint,   
new physics must lie in the phenomena in which  Bose statistics 
is deeply entwined with the structure and dynamics of a liquid. Hence, we must find a starting 
point for disentangling these complicated problems. 
For the shear viscosity, we find such a starting point, in which one can consider one 
problem separately from the other ones.

\begin{figure}
\includegraphics [scale=0.5]{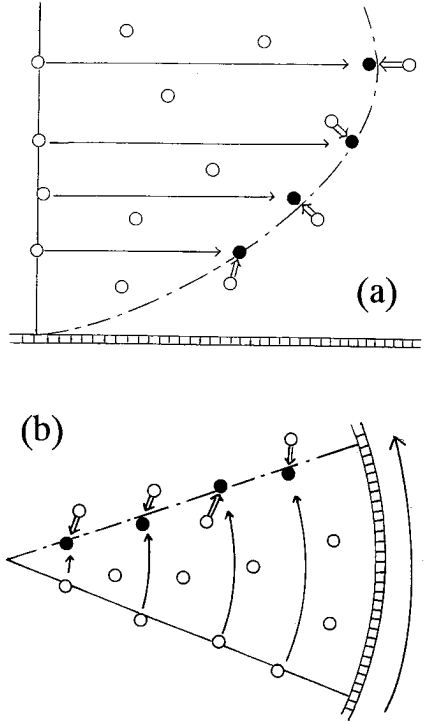}
\caption{\label{fig:epsart} Two types of flows. (a) a flow through a capillary. 
 (b) a rotational flow in a container. 
  White circles represent an initial  distribution of particles. The 
 long arrows move white circles on a solid line to black circles. 
 In both (a) and (b), permutation symmetry holds over the whole 
 liquid at $T<T_{\lambda}$. The particles represented by black circles are reproduced by slight 
 displacements by short arrows from the initial particles close to that 
 particle. }
\end{figure}

 The viscosity has a formally similar expression to other processes of diffusion, such as 
 diffusion of matters and thermal conduction.  The latter are 
 the diffusion of a scalar quantity such as density $\rho$ and  
 temperature $T$, and are always  dissipative processes, whereas the viscosity is  the diffusion of a vector 
 quantity, momentum $\mbox{\boldmath $p$}$.  Since the vector field $m\mbox{\boldmath 
 $v$}(\mbox{\boldmath $r$})$  has various spatial patterns, 
 the viscosity has a qualitatively different feature that is not seen 
 in diffusion of matters and thermal conduction.  Figure 2 
 schematically illustrates two types of flows:(a) a flow in a capillary, and 
 (b) a rotational flow in a container. 
 In fluid dynamics, the viscous dissipation in an incompressible fluid is 
estimated by the dissipation function 
\begin{equation}
	\Phi (\mbox{\boldmath $r$})=2\eta \left(e_{ij}-\frac{1}{3¥}e_{kk}\delta_{ij}\right)^2¥,
	\label{¥}
\end{equation}¥
($e_{ij}=(\partial v_i/\partial x_j+\partial v_j/\partial x_i)/2$ is the 
shear velocity). Particles in a  capillary flow like Fig.2(a) experiences thermal 
dissipation not only at the boundary, but also in a flow. (For  Poiseuille 
flow  Eq.(2), the dissipation function Eq.(3) is not zero at any $r$ except for $r=0$.)
  On the other hand, in a rotating cylindrical container like Fig.2(b), a liquid makes the rigid-body 
 rotation  owing to its viscosity,  the velocity of which is proportional to the radius and the 
 rotational velocity $\Omega$ such as 
 $\mbox{\boldmath $v$}_d( \mbox{\boldmath $r$})  \equiv \mbox{\boldmath 
 $\Omega$}\times \mbox{\boldmath $r$}$ . Except at the boundary  to the wall, there is no 
 frictional force within a liquid, and the flow is therefore a 
 non-dissipative one.  (For  $v_{\phi}(r)=\Omega r$, Eq.(3) is zero at any $r$.)  
  When we discuss the onset mechanism of superfluidity in such a 
 non-dissipative flow, we can take an advantage of this feature to avoid 
 dealing with the dissipation mechanism. 
 
 In this paper, we will consider the frictionless capillary flow of a liquid helium 4 by 
 relating it  with the nonclassical rotational flow using the Kramers-Kronig relation. 
 (The former is superfluidity as a non-equilibrium process, whereas the 
 latter is  superfluidity as a thermal equilibrium one.)
 This method originated in studies of electron 
 superconductivity. Just after the advent of BCS model, an attempt was made to relate the 
 electrical conductivity (more precisely, the micro-wave absorption 
 spectrum) with the penetration depth in the Meissner effect \cite {fer} \cite {ttin}. The former 
 is a quantity in the dissipative phenomenon, and the latter is that in 
 the non-dissipative one. Furthermore, the Meissner effect has a common mechanism 
 with the nonclassical rotational flow of a Bose liquid \cite {noz}. 
 Hence, in the  relationship between the electrical conduction and the 
 Meissner effect in superconductivity, one finds a parallelism to that between the  
 capillary flow and the rotational flow in a liquid helium 4.
  By using this analogy, one can  consider the influence of Bose 
 statistics on the shear viscosity of a liquid without getting involved in details of 
 the dissipation mechanism of a liquid. Furthermore, one can explore the 
 relationship between {\it two typical manifestations of superfluidity \/}.

 Before carrying out this program, one must have a physical image on the 
gradual fall of $\eta $ at $T_{\lambda}<T<2.8K$ observed in a 
liquid helium 4 (Fig.1(a)).  It has been assumed to be a manifestation of the thermal 
fluctuation of Boson systems, and discussed in connection with 
the $\lambda$-shape of the specific heat $C(T)$ (Fig.1(b)). The  
principal peak in $C(T)$ reflects thermal fluctuation in the critical 
region \cite {ahl}.  A peculiar feature of the $\lambda$-shape of $C(T)$ is that  
in addition to the sharp peak, the specific heat displays a symptom of its rise to the 
sharp peak already at $T_{\lambda}<T<T_{\lambda}+0.7K$, which suggests the cause 
of the gradual fall of the shear viscosity in the same temperature region. 
A remarkable fact is that an ideal Bose gas with the same density as the liquid helium 4 shows a 
  similar gradual rise of $C(T)$ around $T_c$ as in Fig.1(c).  
Normally, the enhanced thermal fluctuation at the vicinity of $T_c$  comes from  strong 
competitions between the interaction energy and the entropy effect, and 
 occurs only within the very narrow temperature region around $T_c$. Hence, it is questionable to 
apply the concept of thermal fluctuation to this gradual rise of the 
specific heat at $T_{\lambda}<T<2.8K$, because of the following reasons: (1) it occurs in a far wider 
temperature region than the critical one, and (2) $C(T)$ of an 
ideal Bose gas suggests that it really occurs even if there is no particle 
interactions.  Since the gradual fall of $\eta (T)$ above $T_{\lambda}$ 
may come from the same mechanism as the gradual rise of $C(T)$ above 
$T_{\lambda}$, the thermal fluctuation is a questionable interpretation for both phenomena. 
 (This interpretation of the $\lambda$-shape of $C(T)$ leads us to reconsider the nature of the  
$\lambda$ transition and the meaning of the infinite volume limit. See Sec.5A.)

 Compared to an ordinary liquid, a liquid helium 4 above $T_{\lambda}$ 
 has a  $10^{-3}$ times smaller shear viscosity coefficient.  Although in the normal phase, 
it is already an anomalous liquid  under the strong influence of Bose statistics.
 Hence, it seems  natural to assume that large but not yet macroscopic coherent 
 wave functions gradually grows as a thermal equilibrium state
as  $T\rightarrow T_{\lambda}$ in the normal phase, and 
suppresses the shear viscosity \cite {lod}\cite {mat}. (At an early stage 
in the study of superfluidity, R.Bowers and K.Mendelssohn discussed a 
similar view on the nature of the decrease of $\eta $ above $T_{\lambda}$ \cite {men}.)
The existence of such wave functions must affect the non-dissipative flow as well, such as
 the rotational flow above $T_{\lambda}$. 
 Recently, from this viewpoint, the rotational property of a liquid helium 4  above 
 $T_{\lambda}$ is theoretically examined \cite {koh}, and a slight decrease of 
 the moment of inertia above $T_{\lambda}$ is predicted. (The similar 
 phenomenon occurs in the Meissner effect of the charged Bose gas as well \cite {kohmei}.)  

Experimentally, although a lot of measurements have been done on 
superfluidity of a liquid helium 4, there is so little direct 
experimental information about flow patterns.  Among various 
visualization techniques used in ordinary liquids, {\it particle image velocimetry\/} (PIV), which records 
the motion of micrometre-scale solid particles suspended in the fluid as tracer 
particles, recently becomes available in a superfluid helium 4 \cite {don}\cite {van}. 
Using this technique, we will be able to closely compare the theory and the 
experiment on the onset mechanism of superfluidity.

 In this paper, after formulating the shear viscosity of a liquid, we generalize the result of Ref.14 to the  
 capillary flow.  Since we focus on the continuous change of the system around 
$T_{\lambda}$,  we cannot assume from the beginning the sudden emergence of 
the macroscopic condensate at $T_{\lambda}$.  Rather,  beginning with the 
system without the condensate, we consider the 
Bose system above and below $T_{\lambda}$ on a common ground, and study the intricacy 
underlying the onset of superfluidity.

This paper is organized as follows.   Section 2.A considers the shear viscosity of a 
liquid by relating the dissipative flow to the 
non-dissipative ones using Kramers-kronig relation, and obtains a formula 
describing the influence of Bose statistics on the dissipative flow.
Section.2.B discusses the physical reason that Bose statistics suppresses
the shear viscosity.  Beginning with the Bose system 
{\it without the condensate\/}, Sec.3 formulates a perturbation expansion of the 
susceptibility with respect to the repulsive interaction. By taking 
peculiar diagrams obeying  Bose statistics in the above 
formula, Sec.3 examines how the growth of the coherent wave function 
 gradually suppresses the shear viscosity  at $T>T_{\lambda}$.  
 Section 4 makes a comparison to experiments, and Sec.5 discusses the 
 nature of the $\lambda$ transition, compare the dissipative and the non-dissipative flows, 
 and makes a brief comment on the shear viscosity of Fermi liquids.

\section{Shear viscosity of a Bose liquid}

\subsection{Formalism of the shear viscosity of a liquid}  
  For a liquid helium 4, the  repulsive particle picture is not so unrealistic as it would be 
 for any other liquid. Hence, as its simplest model, we use
\begin{equation}
 H=\sum_{p}\epsilon (p)\Phi_{p}^{\dagger}\Phi_{p}
   +U\sum_{p,p'}\sum_{q}\Phi_{p-q}^{\dagger}\Phi_{p'+q}^{\dagger}\Phi_{p'}\Phi_p , 
   \qquad (U>0),¥¥¥
	\label{¥}
\end{equation}¥
where $\Phi_{p}$ denotes an annihilation operator of a spinless boson. 

In the linear response theory, the shear viscosity coefficient is given by the 
following two-time correlation function of the stress tensor $J_{xy}(t)$
\begin{equation}
   \eta =\frac{1}{Vk_BT¥}¥\int_{0}^{\infty¥}dt<J_{xy}(t)J_{xy}(0)>¥.
	\label{¥}
\end{equation}¥
 In a liquid,  the particle interaction $U$ enhances not only 
 a few-particle or collective excitations but also their relaxations, 
 hence reducing the relaxation time $\tau$, and the shear viscosity coefficient
$\eta =G\tau$. In principle, the shear viscosity coefficient of a liquid must be obtained by 
calculating an infinite series of the perturbation expansion of   
Eq.(5) \cite {kad}.  The interaction $U$ affects  the imaginary part of 
the self energy of particles, thus changing the stress tensor 
$J_{xy}$. To obtain the resulting change of $<J_{xy}(t)J_{xy}(0)>$, 
 one must know the dissipation mechanism of a liquid. In contrast with the case of a gas, the mechanism 
responsible for the shear viscosity in a liquid is similar to the 
vacancy motion in a solid, and is therefore a highly inhomogeneous process in the 
microscopic scale. Hence, it is difficult to derive the shear viscosity 
 of a liquid without using any phenomenological model \cite{jeo}.

Instead of  $\eta $, however, if we apply the linear-response theory  to its 
{\it reciprocal \/} $1/\eta $, and make the perturbation expansion with respect to 
$U$, {\it the increase of $U$ naturally leads to the decrease 
of $\eta$ \/}, thus naturally describing the effect of $U$ on $\eta$.  
Along this line of thought, we regard Eq.(2) of the Poiseuille flow, in 
which $\eta$ appears in the denominator, as a macroscopic linear-response formula.
Without loss of generality, we may focus on a flow velocity at a single point. 
 Using the flow velocity on the axis of rotational symmetry (z-axis),  
we define  a current $\mbox{\boldmath $j$}=\rho\mbox{\boldmath $v$}(r=0)$ on 
the z-axis, and obtains
\begin{equation}
 \mbox{\boldmath $j$}=-\sigma a^2\frac{\Delta \mbox{\boldmath $P$}}{L¥},  \qquad  \sigma =\frac{\rho}{4\eta¥}¥,
	\label{¥}
\end{equation}¥
where $\sigma$ is the conductivity of a liquid in a capillary flow ($\mbox{\boldmath 
$P$}=P\mbox{\boldmath $e$}_z$).  Equation.(6) is a longitudinal 
linear response of a liquid to $\Delta \mbox{\boldmath $P$}/L$, which  includes
the dissipation by the shear viscosity \cite {cur}. The  perturbation 
expansion  of  $\sigma$ with respect to $U$ properly incorporates to $\eta$ the strong-coupling nature 
of a liquid. 

  Let us  generalize  Eq.(6) to the case of the oscillatory pressure gradient as follows
\begin{equation}
 \mbox{\boldmath $j$}(\omega)=-\sigma (\omega) a^2\frac{\Delta \mbox{\boldmath $P$}(\omega)}{L¥}.
	\label{¥}
\end{equation}¥
The Navier-Stokes equation gives an expression of the conductivity spectrum $\sigma (\omega)$ (see Appendix.B).
By definition, $\sigma (\omega)$  in Eq.(7) must contain the mass $m$ of particles.  
 Specifically, $\sigma (\omega)$ satisfies the following sum rule \cite {sum}
\begin{equation}
   \frac{1}{\pi ¥}¥\int_{0}^{\infty¥}\sigma (\omega)d\omega¥=nm¥,
	\label{¥}
\end{equation}¥
where $n$ is the number density of particles.

\begin{figure}
\includegraphics [scale=0.5]{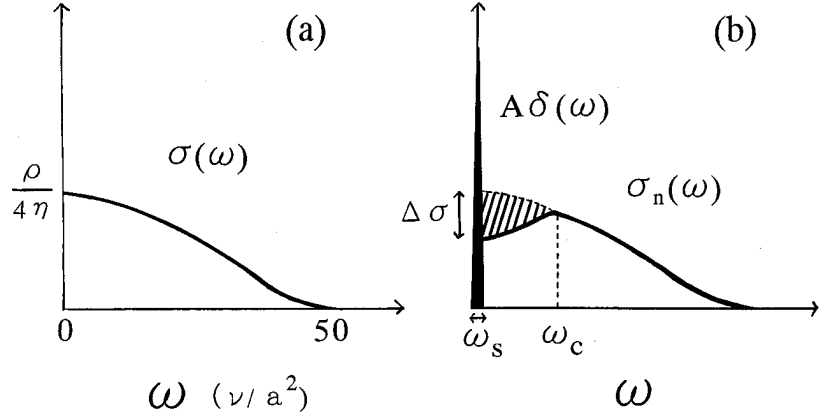}
\caption{\label{fig:epsart} The change of the conductivity spectrum $\sigma (\omega )$ from (a) at 
 $T>T_{\lambda}$ to (b) at $T<T_{\lambda}$.  At $T>T_{\lambda}$,  $\sigma (\omega )$ 
 is given by Eq.(B5).  At $T<T_{\lambda}$, $\omega _s$ is a half width of  
 the sharp peak around \protect$\omega =0$ with an area $A$, and $\omega _c$ 
 is a frequency at the  broad peak in $\sigma _n(\omega )$.  }
\end{figure}

At $T<T_{\lambda}$, a frictionless flow appears in Eq.(7). Characteristic 
to the  frictionless flow is the fact that in addition to  the normal fluid 
part $\sigma _n(\omega)$, the conductivity spectrum $\sigma (\omega)$ has 
a sharp peak around $\omega =0$ (with a very small but finite half width 
$\omega_s$), while satisfying Eq.(8). Hence, one obtains
\begin{equation}
 \mbox{\boldmath $j$}(\omega)=-\left[\sigma _n(\omega)+A\delta(\omega)\right]¥¥a^2 \frac{\Delta \mbox{\boldmath $P$}(\omega)}{L¥},
	\label{¥}
\end{equation}¥
where $\delta(\omega)$ is a simplified expression of the sharp peak around $\omega =0$, 
and $A$ is an area of this peak \cite {fer}.  Figure 3 schematically illustrates such a change of 
$\sigma (\omega)$ when the system passes $T_{\lambda}$.  
When $\sigma (\omega)$ has a form of  $\sigma _n(\omega)+A\delta(\omega)$, 
one obtains the shear viscosity  coefficient $\eta $ in Eq.(6) as
\begin{equation}
   \eta (T)=\frac{1}{4¥}¥\frac{\rho}{\sigma _n(0)+\displaystyle{\left(\frac{A}{\omega _s¥}\right)}¥¥}¥¥.
	\label{¥}
\end{equation}¥
In view of Eq.(10), the sharpness of the peak around $\omega =0$ ($\omega _s \simeq 0$) leads 
to the disappearance of the shear viscosity ($\eta (T) \simeq 0$) at $T<T_{\lambda}$.

Let us generalize  Eq.(7) so that it includes not only the current 
responding in phase with $\Delta \mbox{\boldmath $P$}(\omega)$ but also 
a current responding out of phase.  $\sigma (\omega)$ in Eq.(7) is 
generalized to a complex number $\sigma_1+i\sigma_2$ as follows
\begin{equation}
 \mbox{\boldmath $j$}(\omega)=-\left[\sigma_1(\omega)+i\sigma_2(\omega)\right]¥a^2 \frac{\Delta \mbox{\boldmath $P$}(\omega)}{L¥},
	\label{¥}
\end{equation}¥
where $\sigma (\omega)$ in Eq.(7) is replaced by $\sigma_1(\omega)$. 
Causality states in Eq.(11) that only after the pressure gradient is 
applied, one observes a current, and obtains the following Kramers-Kronig relation
\begin{equation}
   \sigma_1(\omega')=\frac{2}{\pi}¥\int_{0}^{\infty¥}d\omega\frac{\omega\sigma_2(\omega)}{\omega^2-\omega'^2¥}¥,
	\label{¥}
\end{equation}¥
\begin{equation}
   \sigma_2(\omega')=-\frac{2\omega'}{\pi}¥\int_{0}^{\infty¥}d\omega\frac{\sigma_1(\omega)}{\omega^2-\omega'^2¥}¥.
	\label{¥}
\end{equation}¥

We will embed  Eq.(11) in a more general picture, and derive properties of 
a capillary flow by regarding it as a special case. Instead of 
$\Delta \mbox{\boldmath $P$}(\omega)$, we use a quantity directly specifying the 
state of the system, that is, the velocity field $\mbox{\boldmath $v$}(t)$.  Applying the 
 pressure gradient to a liquid is equivalent to assuming the velocity 
field $\mbox{\boldmath $v$}(t)$ satisfying the equation of motion
\begin{equation}
      \frac{d \mbox{\boldmath $v$}(t)}{d t¥}¥= -\frac{\Delta \mbox{\boldmath $P$}(t)}{L¥} ,
	\label{¥}
\end{equation}¥
as an external field \cite {vec}.  
 
  Let us consider a rotational flow like Fig.2(b).  We assume 
 $\mbox{\boldmath $v$}_d(\mbox{\boldmath $r$},t) =\mbox{\boldmath 
 $\Omega$}\times \mbox{\boldmath $r$}e^{i\omega t}$ of the rigid-body 
rotation.  In the rotational flow, $\mbox{\boldmath $j$}(\mbox{\boldmath 
$r$},\omega)$ and $\mbox{\boldmath $v$}_d(\mbox{\boldmath $r$},\omega)$ 
form the perturbation energy like $\int dx\mbox{\boldmath 
$v$}_d\cdot \mbox{\boldmath $j$}$.  Since the rotational flow is a non-dissipative system,   
$\mbox{\boldmath $j$}(\mbox{\boldmath $r$},\omega)$ is a dynamical response of a 
liquid to the mechanical external field $\mbox{\boldmath $v$}_d(\mbox{\boldmath $r$},\omega)$, such as 
$\mbox{\boldmath $j$}(\mbox{\boldmath $r$},\omega)= \chi(\mbox{\boldmath 
$r$},\omega)\mbox{\boldmath $v$}_d(\mbox{\boldmath $r$},\omega)$, where $\chi(\mbox{\boldmath 
$r$},\omega)$ is the susceptibility of the system \cite {noz}.
Replacing $\Delta \mbox{\boldmath $P$}(\omega)/L$ with $-i\omega\mbox{\boldmath $v$}(\omega)$ in Eq.(11), we interchange 
the real and imaginary part of the coefficient in Eq.(11), and make it a local equation as follows 
\begin{equation}
 \mbox{\boldmath $j$}(q,\omega)=\left[-\omega\sigma_2(q,\omega)+i\omega\sigma_1(q,\omega)\right]¥¥a^2
                                                \mbox{\boldmath $v$}(q,\omega).
           \label{¥}
\end{equation}¥
Equation (15) is a general formula including not only $i\omega\sigma_1(q,\omega)$ for the dissipative 
flow (Fig.2(a)), but also $-\omega\sigma_2(q,\omega)$ for the non-dissipative one (Fig.2(b)). 
If one determines $\sigma_2(q,\omega)$ in the non-dissipative flow, 
 one obtains $\sigma_1(q,\omega)$  using Eq.(12) at a given $q$.
 
In the rotational flow, because of $ div\mbox{\boldmath $v$}_d( \mbox{\boldmath $r$},t)=0 $, 
 $\mbox{\boldmath $v$}_d( \mbox{\boldmath $r$},t)$
 acts as a transverse-vector probe to the excitation of bosons. This fact 
 allows us a formal analogy with the vector potential 
 $\mbox{\boldmath $A$}(\mbox{\boldmath $r$})$ in the Coulomb gauge 
 ($div\mbox{\boldmath $A$}(\mbox{\boldmath $r$})=0$) acting on the charged Bose system.   
 Hence, the neutral Bose system has a similar form of current to that of the charged Bose 
 system \cite {noz} such as
\begin{equation}
	J_{\mu}(q,\tau)=\sum_{p,n} 
	\left(p+\frac{q}{2¥}\right)_{\mu}\Phi_p^{\dagger}\Phi_{p+q}e^{-i\omega _n\tau}¥¥¥,
	\label{¥}
\end{equation}¥
($\hbar =1$ and  $\tau =it$). Within the linear response, one obtains the 
generalized susceptibility 
\begin{equation}
     \chi_{\mu\nu}(q,\omega _n)=\frac{1}{V¥}\int_{0}^{\beta¥}d\tau \exp(i\omega_n\tau)
	                          <G|T_{\tau}J_{\mu}(q,\tau)J_{\nu}(q,0)|G>¥¥¥,
\end{equation}¥
 where $|G>$ is the ground sate \cite {flu}.  
Generally, the susceptibility is decomposed into the longitudinal and transverse part ($\mu =x,y,z$)
\begin{equation}
	\chi_{\mu\nu}(q,\omega )=\frac{q_{\mu}q_{\nu}}{q^2¥}\chi^L(q,\omega)     
	            +\left(\delta_{\mu\nu}-\frac{q_{\mu}q_{\nu}}{q^2¥}\right)¥\chi^T(q,\omega) .
	\label{¥}
\end{equation}¥
For the later use, we define a  term proportional to 
 $q_{\mu}q_{\nu}$ in $\chi_{\mu\nu}$ by $\hat{\chi}_{\mu\nu}$
 \begin{eqnarray}
       	\chi_{\mu\nu}(q,\omega)&=&\delta_{\mu\nu}\chi^T(q,\omega)
	                  +q_{\mu}q_{\nu}\left(\frac{\chi^L(q,\omega)-\chi^T(q,\omega)}{q^2¥}\right)¥ \nonumber \\ 
	                          &\equiv& \delta_{\mu\nu}\chi^T(q,\omega)+\hat{\chi}_{\mu\nu}(q,\omega), 
	\label{¥}
\end{eqnarray}¥
where $\hat{\chi}_{\mu\nu}$  represents the balance between the longitudinal and transverse 
susceptibility \cite {wha}. 
In the rotational flow, the influence of the wall motion propagates 
along the radial direction, which is perpendicular to the particle motion 
driven by rotation as illustrated in Fig.2(b). Hence, the rotational 
motion is a transverse response 
described by the transverse susceptibility $\chi^T(q,\omega)$. In the 
right-hand side of Eq.(15), the real part  $-a^2\omega\sigma_2(q,\omega)$ of 
the coefficient describes the rotational motion as a non-dissipative response such as
\begin{equation}
   -a^2\omega\sigma_2(q,\omega)=\chi^T(q,\omega) ¥,
	\label{¥}
\end{equation}¥
and the  imaginary part $a^2\omega\sigma_1(q,\omega)$ describes the dissipation.
 
Let us focus on the capillary flow. Using Eq.(20) in the right-hand side 
of Eq.(12) at a given $q$, one obtains 
\begin{equation}
   a^2\sigma_{1}(q,\omega')= -\frac{2}{\pi}¥¥
         \int_{0}^{\infty¥}d\omega\frac{\chi^T(q,\omega)}{\omega^2-\omega'^2¥}¥.
	\label{¥}
\end{equation}¥
The slow excitations ($\omega\simeq 0$) in $\chi^T(q,\omega)$ of Eq.(21) 
are important for determining $\sigma_{1}(q,\omega)$. On this point, Eq.(21) has different 
interpretations in the normal and the superfluid phase.  In the normal 
fluid phase, $\chi^L(q,\omega)=\chi^T(q,\omega)$ is satisfied at small $q$ and $\omega$, 
and one can replace $\chi^T(q,\omega)$ in Eq.(21) by $\chi^L(q,\omega)$
 at a small $q$ and $\omega$ \cite {def}.  Hence, the conductivity $\sigma_{1n}(q,\omega)$ 
 of a capillary flow in the normal fluid phase is given by 
\begin{equation}
   a^2\sigma_{1n}(q,\omega')= -\frac{2}{\pi}¥¥
         \int_{0}^{\infty¥}d\omega\frac{\chi^L(q,\omega)}{\omega^2-\omega'^2¥}¥ .
	\label{¥}
\end{equation}¥

In the superfluid phase, under the strong influence of Bose statistics, 
the condition of $\chi^L(q,\omega)=\chi^T(q,\omega)$  at $q\rightarrow 
0$ is violated in the low-energy region extending from  $\omega$ =0 (see Sec.2.B). 
(Furthermore, the rigidity of superfluidity requires that the nonzero 
$\chi^L(q,\omega)-\chi^T(q,\omega)$ depends on $q$ and $\omega$ very weakly at  small $q$ and $\omega$.) 
Consequently, one cannot replace  $\chi^T(q,\omega)$ by 
$\chi^L(q,\omega)$ in Eq.(21). For $\sigma_1(q,\omega)$, in addition to $\sigma_{1n}(q,\omega)$, 
one must separately consider the contribution of $\chi^T(q,\omega)-\chi^L(q,\omega)$. 
For $\sigma_1(0,\omega)$ of the capillary flow, one obtains  
\begin{equation}
   \sigma_1(\omega')=\sigma_{1n}(\omega')+
            \frac{2}{\pi a^2¥}¥\int_{0}^{\infty¥}d\omega
            \frac{\lim_{q\to 0}[\chi^L(q,\omega)-\chi^T(q,\omega)]¥}{\omega^2-\omega'^2¥} ¥,
	\label{¥}
\end{equation}¥
 where  we replace $\chi^L(q,\omega)-\chi^T(q,\omega)$ by its limit at $q\rightarrow 0$.
Since this limit depends on $\omega$ very weakly at small $\omega$, 
we take out it from the integral with the aid of
 \begin{equation}
  \int_{0}^{\infty¥}¥\frac{d\omega}{\omega^2-\omega'^2¥}¥=\delta(\omega '),
	\label{¥}
\end{equation}¥
and obtains
\begin{equation}
   \sigma_1(\omega)=\sigma_{1n}(\omega)+\frac{2}{\pi a^2¥}¥\lim_{q\to 0}[\chi^L(q,0)-\chi^T(q,0)]\delta(\omega). ¥
	\label{¥}
\end{equation}¥
Comparing Eq.(25) with Eq.(9), one obtains an expected form of $\sigma(\omega)$ for the superfluid flow 
with $A=2\lim_{q\to 0}[\chi^L(q,0)-\chi^T(q,0)]/(\pi a^2)$. If 
superfluidity was perfectly rigid to any perturbations, $\lim_{q\to 
0}[\chi^L(q,\omega)-\chi^T(q,\omega)]$ would have no frequency dependence. In reality,
 this limit weakly depends on $\omega$ in Eq.(23). Hence, although the 
 peak is well approximated by $\delta(\omega)$ in Eq.(25), it has  
a small but finite half width $\omega _s$ as in Fig.3(b), which represents a 
degree of the rigidity  of superfluidity.

The change of $\sigma(\omega)$ from Fig.3(a) to 3(b) occurs under the sum 
rule Eq.(8).  The area $A$ of the sharp peak is equal to that of the 
shaded region in Fig.3(b).  We approximate this shaded region by a triangle. (A broad peak 
 of $\sigma_{n}(\omega)$ is located at $\omega_c$). At $\omega =0$, the 
 conductivity is enhanced by the sharp peak as $A/\omega _s$, 
but the original $\sigma (0)$ changes to $\sigma (0)-\Delta\sigma $ by 
the sum rule. We take into account the sum rule as $-\Delta\sigma  
\omega_c/2+A=0$, hence $\Delta\sigma =2A/\omega_c$. 
After all, the conductivity at $\omega =0$ becomes $[\sigma (0)-2A/\omega_c]+A/\omega_s$ 
at $T<T_{\lambda}$. Using this form and the expression of $A$ in Eq.(25), one obtains $\eta =\rho/(4\sigma (0))$ as
\begin{equation}
  \eta (T)=\left(\frac{\rho}{4¥}\right)¥\frac{1}{\sigma_{1n}(0)
              +\displaystyle{\frac{2}{\pi a^2¥}\left(\frac{1}{\omega_s¥}-\frac{2}{\omega_c¥}¥\right)
                                                             \lim_{q\to 0}[\chi^L(q,0)-\chi^T(q,0)]¥¥}}¥.
	\label{¥}
\end{equation}¥
Here, we define {\it the mechanical superfluid density \/} 
 $\hat {\rho _s}(T)$ ($\equiv \lim_{q\to 0}[\chi^L(q,0)-\chi^T(q,0)]= \lim_{q\to 0}[q^2/(q_{\mu}q_{\nu})] \hat {\chi} _{\mu\nu}$), 
  which does not always agree with the conventional thermodynamical superfluid density 
$\rho _s(T)$.  (By ``thermodynamical'', we imply the quantity that 
remains finite in the $V\rightarrow \infty $ limit.)  Using  $\omega_s¥ \ll 
\omega_c$ in Eq.(26), we obtain the following formula of {\it the kinematic shear viscosity\/} 
$\nu(T)=\eta(T)/\rho(T)$ 
\begin{equation}
  \nu (T)=¥\frac{\nu _n}{1
              +\displaystyle{\frac{8}{\pi a^2\omega_s¥}\frac{\hat {\rho _s}(T)}{\rho}}\nu _n¥¥}¥,
	\label{¥}
\end{equation}¥
where $\nu _n$ is the kinematic shear viscosity in the normal fluid 
phase, and it satisfies $\sigma_{1n}(0)=1/(4\nu _n)$ in Eq.(26).  

One notes the following features in the formula (27).

(1) $\nu (T)$ of a superfluid is expressed as an infinite power series of  
$\nu _n$, and the influence of Bose statistics appears in its 
coefficients. This result does not depend on the specific model of a 
liquid, but on the general argument (see Sec.5.B). (The microscopic 
derivation of $\nu _n$ is a subject of the liquid theory.)

(2) Because of $\omega _s\simeq 0$, which characterizes the sharp peak 
in Fig.3(b), a small change of $\hat {\rho _s}(T)$ in the 
denominator of the right-hand side is strongly enhanced to an observable change of $\nu (T)$. 

(3) The existence of $1/a^2$ in front of $\hat {\rho _s}(T)/\rho$ 
indicates that a frictionless superfluid flow appears 
only in a narrow capillary with a small radius $a$: A narrower capillary shows a cleaner evidence 
of a frictionless flow.

We can simply examine the condition of $\hat {\rho _s}(T)\ne 0$ by ignoring the repulsive 
interaction. Using Eq.(16) in Eq.(17), one obtains the following   
term proportional to $q_{\mu}q_{\nu}$ in $\chi _{\mu\nu}$ 
\begin{equation}
\hat{\chi}_{\mu\nu}(q,\omega)  
	            =-\frac{q_{\mu}q_{\nu}}{4¥}¥
	                          \frac{1}{V¥}\sum_{p}\frac{f(\epsilon (p))-f(\epsilon (p+q))}
	                                       {\omega+\epsilon (p)-\epsilon (p+q)¥}¥,
	\label{¥}
\end{equation}¥
where $f(\epsilon (p))$ is the Bose distribution.
When bosons would form the condensate, $f(\epsilon (p))$ in Eq.(28) is a macroscopic 
number for $p=0$ and nearly zero for $p\ne 0$. Thus, in the sum over $p$ in the 
right-hand side of Eq.(28), only two terms corresponding to $p=0$ and 
$p=-q$ remain, with a result that 
\begin{equation}
	\hat{\chi}_{\mu\nu}(q,0)=mn_c(T)¥\frac{q_{\mu}q_{\nu}}{q^2¥}¥,
\end{equation}¥ 
where $mn_c(T)$ is the thermodynamical superfluid density $\rho _s(T)$, and $n_c(T)$ is the 
number density of particles participating in the condensate.  In view of 
Eq.(29), we find that the mechanical superfluid density $\hat 
{\rho _s}(T)$ in this case is equal to the thermodynamical superfluid 
density $\rho _s(T)$. 

 When bosons form no condensate, the sum over $p$ in Eq.(28) is carried out 
 by replacing it with an integral, and one notes
that $q^{-2}$ dependence disappears in the result. Because of $\hat {\rho _s}(T)=0$, 
 one obtains $\nu (T)=\nu _n$ in Eq.(27). This means that, without the interaction between particles, 
BEC is the necessary condition for the superfluid flow.
 Equation.(27) shows that unless the particle interaction is 
taken into account, a superfluid flow abruptly appears at $T_{\lambda}$. 
To explain the gradual decrease of the shear viscosity 
just above $T_{\lambda}$ as in Fig.1(a), we must obtain  
$\chi^T(q,\omega)-\chi^L(q,\omega)$ under the repulsive interaction $U$.

\subsection{Bose statistics and repulsive interaction}

There is a physical reason to expect that the shear viscosity of a Bose 
liquid falls off at low temperature. 
 Quantum mechanics states that in the decay from an excited state with an 
 energy level $E$ to a ground state with $E_0$, the higher excitation 
 energy $E$ gives rise to the shorter relaxation time $\tau $. The 
 time-dependent perturbation theory gives us
 \begin{equation}
 	\frac{\hbar}{\tau ¥}¥ \simeq |E-E_0|.
 	\label{¥}
 \end{equation}¥
  In Eq.(21), $\sigma _{1}(q,\omega) =\rho /(4G\tau )$ in the left-hand 
  side includes the relaxation time $\tau $, whereas 
 $\chi^T(q,\omega)$ in the right-hand side  includes the 
excitation spectrum, hence the difference of energy  $|E-E_0|$.  
 In this sense,  Eq.(21) is a many-body theoretical expression of 
Eq.(30). Microscopically, the structural characteristic of liquids is the  
 {\it irregular arrangements \/} of their molecules. These 
arrangements are similar, and therefore their energies differ only slightly 
from one arrangement to another. The transformation from one arrangement 
to another with a small $|E-E_0|$ 
continuously occurs with a large $\tau $.  Since the fall of the 
shear viscosity at the vicinity of $T_{\lambda}$ is attributed to the decrease of 
$\tau $, Eq.(30) says that the excitation energy $E$ must rise owing to 
Bose statistics. The relationship between 
the excitation energy and  Bose statistics dates back to 
Feynman's argument on the scarcity of the low-energy excitation in a liquid helium 
4 \cite{fey1}, in which he explained  how Bose 
statistics affects the many-body wave function in configuration space. 
To the shear viscosity, we will apply his explanation.

Consider the velocity field in Fig.2(a) (depicted by long thin arrows) that moves 
white circles on a solid straight line to black circles on a 
one-point-dotted-line curve. (A viscid liquid would show such a spatial 
gradient of the velocity field. The influence of adjacent layers in a flow propagates along the 
direction perpendicular to the particle motion. Hence, the excitation 
caused by the shear viscosity is a transverse excitation.) 
Let us assume that a liquid in Fig.2(a) is in the BEC phase, and 
the many-body wave function has permutation symmetry everywhere in a capillary.  
At first sight, these displacements by long arrows seem to be a large-scale configuration change,  
 but this result is reproduced by a set of slight displacements by short 
 thick arrows. In contrast with the longitudinal 
displacements, the transverse displacement does not change the particle 
density in the large scale, and therefore it is always possible to find, 
in the initial configuration,  a particle being close 
to the particle after displacement \cite {log}. 
  In Bose statistics, owing to permutation symmetry, one cannot 
distinguish between two types of particles after displacement, one  moved 
from the neighboring position by the short arrow, and the other moved 
from distant initial positions by the long arrow. {\it Even if  the 
displacement made by the long arrows is a large 
displacement in classical statistics, it is only a slight 
displacement  by the short arrows in Bose statistics \/}. 

To grasp the whole situation, it is useful to imagine this situation in the 
3N-dimensional configuration space. The excited state made by short 
arrows lies in a small distance from the ground state in configuration space. 
Since the excited state is orthogonal to the ground state in the configuration integral,
 the many-body wave function corresponding to the excited state must 
spatially oscillate. Accordingly, it oscillates within a small 
distance in configuration space. Since the kinetic energy of the system 
is determined by the 3N-dimensional gradient of the many-body wave function, this 
steep rise and fall of the amplitude implies that the  excitation energy is 
not small. 

 In coordinate space, although the existence of irregular 
arrangements of molecules, whose energies only slightly differ from that of the ground 
state, is a structural characteristic of a liquid, the above result means 
that its number remarkably decreases due to Bose statistics, leaving only excited states with not low energies. 
The relaxation from such an excited state is a rapid stabilization process   
with a small $\tau $. This mechanism explains why  Bose statistics reduces the shear viscosity 
coefficient $\eta =G\tau $ \cite {ord}. Formally, while the longitudinal 
excitations are ensured by the particle conservation to satisfy 
$\chi^L(q,0)=\rho$ in Eq.(25) both above and below $T_{\lambda}$, the low-energy transverse 
excitations become scarce below $T_{\lambda}$, and it reduces $\chi^T(q,0)$ at $q\rightarrow 0$. 
Hence, $\sigma _1(\omega)$ has a $\delta$-function form 
 in Eq.(25) and $\hat {\rho _s}(T)$ becomes nonzero, thus leading to $\nu (T)\rightarrow 0$ in Eq.(27).

In Sec.2A, we considered the rotational flow as a complementary phenomenon of the 
capillary flow. As discussed in the argument above Eq.(20), the particle motion by rotation is 
a transverse excitation. In Fig.2(b), the similar argument holds on the  
role of Bose statistics: Two types of particles, one from distant place 
by the long curved arrow, and 
the other from neighboring place by the short arrow are indistinguishable, hence leading to 
the scarcity of the low-energy transverse excitation as discussed in Ref.14. 
 Under the slow rotation, the region near the center of rotation 
 decouples from the motion of a container \cite {hes}\cite {pac}. 
A common mechanism lies in the two typical manifestations of superfluidity, 
the frictionless flow in a capillary and the decrease of the moment of 
inertia in the  rotational flow in a container.

 When the system is at high temperature ($\mu \ll 0$), the coherent wave function has a microscopic size.  
 If a long arrow in Fig.2 takes a particle to a position beyond the coherent wave function 
 including that particle, one cannot regard the particle after 
 displacement as an equivalent of the initial one.  The mechanism below 
$T_{\lambda}$ in  Fig.2 does not work for the large  displacement extending over two different wave 
functions. Hence, we obtain $\chi^T(q,0)=\chi^L(q,0)$ at $q\rightarrow 0$ 
in Eq.(25), and $\nu (T)=\nu _n$ in Eq.(27).

\begin{figure}
\includegraphics [scale=0.5]{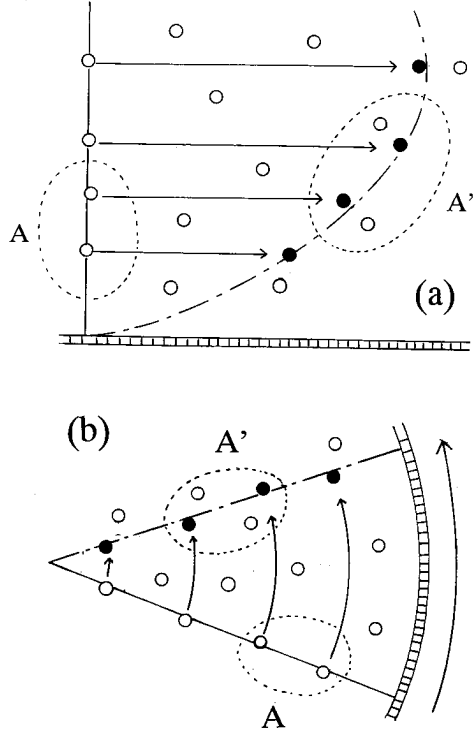}
\caption{\label{fig:epsart} Two types of flows just above $T_{\lambda}$. Areas enclosed by 
 a dotted line represent large but not yet macroscopic size of coherent 
 wave functions, a size of which is exaggerated for clarity.  }
\end{figure}

From our viewpoint, a remarkable state of the boson system lies at the 
vicinity of $T_{\lambda}$ in the normal phase ($\mu -\Sigma \leq 0$).  In 
Fig.4, the coherent many-body wave function grows to a large but not 
yet a macroscopic size.  (Approximately, the chemical potential $\mu$, hence $\mu 
-\Sigma$ as well, is inversely proportional to  
 the size of the coherent many-body wave function \cite{mat}, which 
 corresponds to the size of regions enclosed by a dotted line in Fig.4.) 
 The permutation symmetry holds only within each of these regions.  
 When particles are moved from a region A 
to another region A' in Fig.4(a), the mechanism below $T_{\lambda}$ does not work. 
But the repulsive interaction $U$, not only determining $\nu _n$ in Eq.(27), 
but also affects the role of Bose statistics as follows.  
 In general, when  a particle moves in the interacting system, 
it induces the motions of other particles. In particular, the
long-distance displacement of a particle in coordinate space causes the 
excitation of many  particles, and therefore it needs a high excitation 
 energy.  Hence,  the short-distance displacement is a major 
 ingredient of the low-energy excitation.  
 This means that {\it excited particles are not likely to go beyond  a single 
coherent wave function, but likely to remain in it. \/} Hence, under the 
repulsive interaction, the suppression mechanism of the low-energy 
excitation owing to Bose statistics works just above $T_{\lambda}$ as well.  
 (A similar argument holds in the rotational flow in Fig.4(b) \cite {koh}.)

This interplay between Bose statistics and the repulsive interaction will 
appear as follows. If we increase the strength of $U$, the excited 
bosons get to remain in the same coherent wave function, and therefore 
the low-energy transverse excitation will raises its energy. The relaxation time to 
 the ground state becomes short, thus making the shear viscosity fall off. 
In Eq.(26), the condition of $\chi^L(q,0)=\chi^T(q,0) $ at 
$q\rightarrow 0$ will be violated at a certain critical value of $U$. In 
real materials, we realize this situation in an alternative manner. 
 If we decrease the temperature at a given $U$, the coherent wave 
function obeying Bose statistics grows. The above condition will be violated 
at {\it a certain temperature \/}, which will give the onset 
temperature of the decease of $\eta $ in Fig.1(a).

\section{Susceptibility and shear viscosity above $T_{\lambda}$}
 Let us formulate the influence of Bose statistics on the shear 
 viscosity.  Equation (26) says that $\chi^L(q,\omega)-\chi^T(q,\omega)$ is a crucial quantity for this 
 purpose. The problem is how to formulate in this quantity the interplay between Bose 
 statistics and the repulsive interaction.  Under 
 the repulsive interaction $\hat{H_I}(\tau)$, one considers the integrand of Eq.(17) as
  \begin{equation}
		<G|T_{\tau}J_{\mu}(x,\tau)J_{\nu}(0,0)|G>  
       =  \frac{\displaystyle{<0|T_{\tau}\hat{J}_{\mu}(x,\tau)\hat{J}_{\nu}(0,0)
 	              exp\left[-\int_{0}^{\beta¥}d\tau \hat{H}_I(\tau)¥\right]|0>¥}}
 	        {\displaystyle{<0|exp\left[-\int_{0}^{\beta¥}d\tau  \hat{H}_I(\tau)¥\right]|0>¥}}¥.
	\label{¥}
\end{equation}¥

\begin{figure}
\includegraphics [scale=0.5]{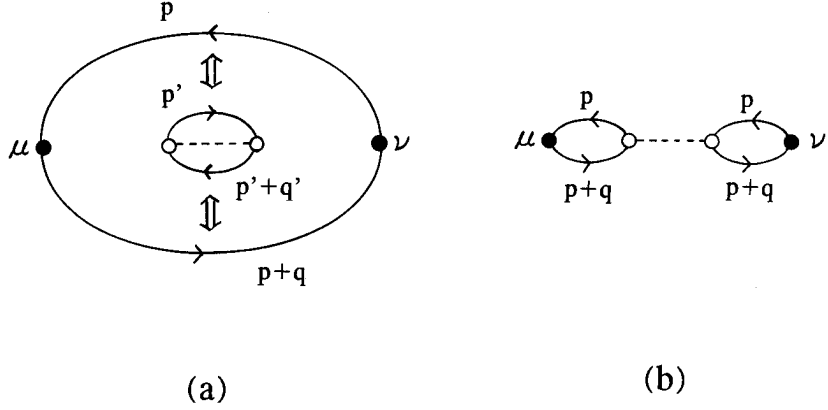}
\caption{\label{fig:epsart} The process of line exchanges due to Bose statistics 
         occurring in $J_{\mu }J_{\nu }exp[-\int d\tau H_I(\tau )]$ of Eq.(31).
         Two times of the line exchange in Fig.5(a), between a current-current 
           response tensor $J_{\mu}J_{\nu}$ (a large bubble) and a bubble  
           excitation by the repulsive interaction $U$ (an 
           inner small bubble with a dotted line), yields Fig.5(b). 
        (The black and white small circle represents a vector and scalar vertex, respectively.)   }
\end{figure}

 Figure 5(a) illustrates  the current-current response tensor 
 $\hat{J}_{\mu}(x,\tau)\hat{J}_{\nu}(0,0)$ (a large bubble with $\mu $ 
 and $\nu $) in the medium:  
 Owing to $exp(-\smallint \hat{H}_I(\tau)d\tau)$ in Eq.(31), 
 scatterings of particles frequently occur in $|G>$ as illustrated by an 
 inner small bubble with a dotted line $U$.  (The black 
and white circle represents a coupling to $\mbox{\boldmath $v$}_d(\mbox{\boldmath $q$})$
 and $U$, respectively.)  The state $|G>$ includes many interaction 
 bubbles like the small one with various momentums $p'$ and $p'+q'$.
 A solid line with an arrow represents
\begin{equation}
G(i\omega _n,p)=\frac{1}{i\omega _n-\epsilon(p)-\Sigma+\mu¥}.
	\label{¥}
\end{equation}¥
($\mu$  is a chemical potential implicitly determined by 
$V^{-1}\Sigma[\exp(\beta[\epsilon (q)+\Sigma-\mu])-1]^{-1}=n$).
Owing to the repulsive interaction, the boson has a self energy 
 $\Sigma$ ($>0$) (we ignore its $\omega$ and $p$ dependence by assuming it 
 small). With decreasing temperature, the negative $\mu$ at 
 high temperature approaches a small positive value of $\Sigma$, finally 
 reaching   Bose-Einstein condensation satisfying $\mu=\Sigma$.

When the system is just above $T_{\lambda}$ in the normal phase,  
particles in the ground state  $|G>$ experience 
the strong influence of Bose statistics. Hence, the perturbation 
of $\hat{J}_{\mu}(x,\tau)\hat{J}_{\nu}(0,0)$ in 
Eq.(31) must be developed in such a way that, as the  order of the perturbation increases, 
it gradually include a new effect due to Bose statistics.
 Specifically, the large bubble $\hat{J}_{\mu}(x,\tau)\hat{J}_{\nu}(0,0)$ and the small bubble 
  in Fig.5(a) form a coherent wave function as a whole. 
  When one of the two particles in the large  bubble and in the small bubble have the same momentum 
($p=p'$), and the other in both bubbles have another   
 same momentum ($p+q=p'+q'$) in Fig.5(a), a graph made by 
exchanging these particle lines must be included in the expansion. 
The exchange of two lines having $p$ and $p'$($=p$), and the exchange of 
the other two lines having $p+q$ and $p+q'$($=p+q$) yield Fig.5(b). 
The result is that two bubbles with the same momentums are linked by the 
repulsive interaction, yielding the following contribution to $\chi_{\mu\nu}(q,\omega)$ 
\begin{equation}
  U\frac{1}{V¥}\sum_{p}(p+\frac{q}{2¥}¥)_{\mu}(p+\frac{q}{2¥}¥)_{\nu}
	                 \left[-\frac{f(\epsilon (p)+\Sigma)-f(\epsilon (p+q)+\Sigma)}
	                            {\omega+\epsilon (p)-\epsilon (p+q)¥} \right]^2¥¥¥.
	\label{¥}
\end{equation}¥

(a) With decreasing temperature, the coherent wave function grows to a large 
size, and the exchange of particle lines owing to Bose statistics like Fig.5 occurs 
many times. Hence, one cannot ignore the higher-order terms in Eq.(31), 
which become more significant with the growth of the coherent wave 
function. Continuing these exchanges, one obtains the $l$-th order term 
which has two vector vertex (black circles) at both ends, to which 
$(p+\frac{q}{2¥}¥)_{\mu}(p+\frac{q}{2¥}¥)_{\nu}$ is attached, 
and $2l$ scalar vertex (white circles) between the two ends. 

(b) Among physical processes in Eq.(33), it becomes evident with 
decreasing temperature that a process including zero momentum particles plays a dominant role.  
Specifically, a process with $p=0$ corresponds to an excitation from the rest 
particle, and a process with $p=-q$ corresponds to a decay into the rest one. 
These processes are important in the higher-order terms as well. 

With these considerations (a) and (b), we obtain the following form of 
$\hat{\chi}_{\mu\nu}(q,0)=[\chi^L(q,0)-\chi^T(q,0)]q_{\mu}q_{\nu}/q^2$ at the vicinity of $T_{\lambda}$
\begin{equation}
	\hat{\chi}_{\mu\nu}(q,0)=\frac{q_{\mu}q_{\nu}}{2¥}¥\frac{1}{V¥}\sum_{l=0}^{\infty¥}U^lF_{\beta}(q)^{l+1},
	\label{¥}
\end{equation}¥
where
\begin{equation}
	 F_{\beta}(q)= \frac{(\exp(\beta[\Sigma-\mu]-1))^{-1}-(\exp(\beta[\epsilon (q)+\Sigma-\mu])-1)^{-1}} 
	                         {\epsilon (q)¥¥} ¥
	\label{¥}
\end{equation}¥
 is a positive monotonously decreasing function of $q^2$, which approaches zero as 
$q^2\rightarrow \infty$.  

At a high temperature ($\beta\mu \ll 0$), $F_{\beta}(q)$ is 
small, and it guarantees the convergence of an infinite series in Eq.(34), with a result that
\begin{equation}
	\hat{\chi}_{\mu\nu}(q,0)=\frac{q_{\mu}q_{\nu}}{2¥}¥
	                 \frac{1}{V¥}\frac{F_{\beta}(q)}{1-UF_{\beta}(q)¥}.¥¥
	\label{¥}
\end{equation}¥
With decreasing temperature, however, the negative $\mu$ gradually 
approaches $\Sigma$, hence  $\Sigma-\mu\rightarrow 0$. Since  
$F_{\beta}(q)$ increases as $\Sigma-\mu\rightarrow 0$, it   
makes the higher-order term significant in Eq.(34). 
 Since $F_{\beta}(q)$ is a positive decreasing function of $q^2$, the divergence of Eq.(36)
 first occurs at $q^2=0$. An expansion of $F_{\beta}(q)$ around $q^2=0$,
 $F_{\beta}(0)-bq^2+\cdots$ has a form such as
\begin{equation}
	 F_{\beta}(q) =\frac{\beta}{4\sinh ^2 \displaystyle{\left(\frac{|\beta[\mu(T)-\Sigma]}{2¥}\right)}¥¥¥}
	                   \left[1-\frac{\beta}{2¥}\frac{1}{\tanh  \displaystyle{\left(\frac{|\beta[\mu(T)-\Sigma]|}{2¥}¥\right)}¥¥}
	                        \frac{q^2}{2m¥}¥¥¥¥  +\cdots    \right]¥¥  .
	\label{¥}
\end{equation}¥
 At $q\rightarrow 0$, the denominator $1-UF_{\beta}(q)¥$ in the 
 right-hand side of Eq.(36) 
has a form of $[1-UF_{\beta}(0)]+Ubq^2¥$.  With decreasing temperature ($\Sigma -\mu\rightarrow 0$),
 $UF_{\beta}(0)$ increases and finally reaches 1, that is,  
\begin{equation}
     U\beta=4\sinh ^2\left(\frac{\beta[\mu (T)-\Sigma(U)]}{2¥}¥\right)¥ .
	\label{¥}
\end{equation}¥
At this point, the denominator in Eq.(36) gets to begin with $q^2$, and 
$\hat{\chi}_{\mu\nu}(q,0)$ therefore changes to have a form of 
$q_{\mu}q_{\nu}/q^2$  at $q\rightarrow 0$, and $q_{\mu}q_{\nu}/q^2$ has  a 
non-zero coefficient $F_{\beta}(0)/(2VUb)$. In view of the definition of 
Eq.(19), this result implies  $\lim_{q\to 0}[\chi^L(q,0)-\chi^T(q,0)]\ne 0$ in Eq.(26). 
 From now, we call $T$ satisfying Eq.(38)  {\it the onset temperature \/}  $T_{on}$ \cite {com}. 

At $T=T_{on}$, substituting Eq.(37) into Eq.(36), we obtain $\hat{\chi}_{\mu\nu}$ at $q\rightarrow 0$
 \begin{equation}
	\hat{\chi}_{\mu\nu}(q,0)=\frac{2m}{U\beta_{on}¥}\frac{1}{V¥}
	                   \tanh \left(\frac{|\beta _{on}[\mu(T_{on})-\Sigma]|}{2¥}¥\right)¥¥
	                              \frac{q_{\mu}q_{\nu}}{q^2¥}¥ .
\end{equation}¥
 By the definition of $\hat {\rho _s}(T)=\lim _{q\to 
 0}(q^2/q_{\mu}q_{\nu}) \hat {\chi} _{\mu\nu}$, one obtains with the aid of Eq.(38) 
\begin{equation}
	\hat {\rho _s}(T_{on})=\frac{1}{V¥}\frac{m}{\sinh |\beta _{on}[\mu(T_{on})-\Sigma] |¥}¥.
\end{equation}¥
The {\it  mechanical superfluid density\/} $\hat {\rho _s}(T)$ is given by $mc(T)n_0(T)$,  where 
\begin{equation}
	c(T)=\frac{2}{\exp(\beta |\mu(T)-\Sigma|)+1¥}¥
\end{equation}¥
 is a  Fermi-distribution-like coefficient, and 
\begin{equation}
     n_0(T)=\frac{1}{V¥}\frac{1}{\exp(-\beta [\mu(T)-\Sigma])-1¥}¥¥
	\label{¥}
\end{equation}¥
is the number density of $p=0$ bosons. 
In the $V\rightarrow\infty $ limit, this quantity is normally 
regarded to be zero. For the finite system just above $T_{\lambda}$,  however,
 $n_0(T)$ has a large but not yet macroscopic value. In real finite system, its magnitude must 
be estimated by experiments (see Sec.4).

The violation of $\chi^T(q,0)= \chi^L(q,0)$ at $q\rightarrow 0$ gives 
rise to various anomalous mechanical responses of a superfluid. At the 
vicinity of $T_{\lambda}$, Eq.(38) is approximated as 
$U\beta=\beta^2[\mu (T)-\Sigma(U)]^2$ for a small $\mu-\Sigma$. This 
 condition has two solutions $\mu (T)=\Sigma(U)\pm \sqrt{Uk_BT}$. It is 
 generally assumed that the repulsive Bose system undergoes BEC 
  as well as a free Bose gas. Hence, with decreasing temperature, 
  $\mu (T)$ of repulsive Bose system should reach $\Sigma (U)$ at a finite temperature, 
  during which course the system necessarily passes a 
 state satisfying  $\mu (T)=\Sigma (U)-\sqrt{Uk_BT}¥$. 
 Consequently, the anomalous mechanical response of a superfluid appears 
 prior to $T_{\lambda}$. 
 Using Eq.(40) in the formula of {\it the kinematic shear viscosity \/} 
 (Eq.(27)), we obtain $\nu (T)$
 at the vicinity of $T_{\lambda}$, which {\it  always falls off prior 
 to BEC \/}, that is, $T_{on}>T_{\lambda}$. Among various mechanical responses of a 
 liquid helium 4 above $T_{\lambda}$, the shear viscosity coefficient most 
 evidently shows a change toward superfluidity far above the critical region. 
 This feature comes from the structure of Eq.(27), in which the small $\omega_s$ amplifies a  
 small $\hat{\rho _s}(T)/\rho$ to an observable change of $\nu (T)$. 
 
 The phenomenon at $T_{on}$ has the following difference from thermal fluctuations. 
Since the thermal fluctuation is essentially a local phenomenon, 
 its influence already appears in the low-order terms of the perturbation 
expansion in which only a few particles participate. 
In Eq.(34), however, only after the perturbation expansion is summed up to the infinite 
order, the non-zero $\hat {\rho _s}(T)$ defined as 
$\lim_{q\to 0}[q^2/(q_{\mu}q_{\nu})] \hat {\chi} _{\mu\nu}$ appears \cite {lod} \cite {pha}.
 
 With decreasing temperature from $T_{on}$, in addition to the particle 
 with $p=0$, other particles having small but 
 finite momentums get to contribute to the $1/q^2$ divergence of 
 $\hat{\chi}_{\mu\nu}(q,0)$ as well. (In addition to Eq.(35), a new $F_{\beta}(q)$ 
 including $p \ne 0$  also satisfies $1-UF_{\beta}(0)=0$ in Eq.(36).) 
 Hence,  {\it the mechanical superfluid density\/} increases.  
 On this change, we have the following physical explanation. 
 For the  repulsive Bose system, particles are likely to spread 
uniformly in coordinate space due to the repulsive force. This feature 
makes the particles with $p\ne 0 $ behave similarly with other particles, 
especially with the particle having zero momentum.
 {\it If they behave differently from others, a resulting 
locally high density of particle raises the interaction energy.\/}  This is a 
reason why {\it many particles participate in the singular dynamical behavior
even at a temperature in which only a few particles participate in the 
coherent many-body wave function \/}. In contrast with the 
thermodynamical superfluid density $\rho _s(T)$, the mechanical superfluid density 
$\hat{\rho _s}(T)$ is a concept including such a dynamics of the system. 
 
When the  system reaches $T=T_{\lambda}$ in which $\mu=\Sigma$ is 
satisfied, one notes $c(T_{\lambda})=1$ and $n_0(T_{\lambda})=n_c$. This means that the zero 
momentum part of $\hat{\rho _s}(T_{\lambda})$ agrees with the 
conventional thermodynamical superfluid density $\rho 
_s(T_{\lambda})=mn_c$, and $n_c$ abruptly grows to a macroscopic number. While   
$\hat{\rho _s}(T)$ at $T>T_{\lambda}$ vanishes in the $V\rightarrow \infty$ limit, 
$\hat{\rho _s}(T)$ at $T<T_{\lambda}$ remains finite in this limit. 
Hence, in thermodynamics, only the latter remains.  In the mechanical 
response of the finite system, however, $\hat{\rho _s}(T)$ manifests itself already at $T>T_{\lambda}$.

\section{Comparison to experiments}
 At $T>2.8K$ in 1 atm, the shear viscosity of a liquid helium 4
 slightly increases with decreasing temperature.  As shown in Fig.1(a), after reaching a 
 maximum value at 2.8K, it begins to reduce its value. On the contrary, 
 the shear viscosity of a classical liquid has the general property of 
 increasing monotonously with decreasing temperature \cite {she}. 
At first sight, a liquid helium 4 seems to undergo a crossover  
 from a classical to a quantum liquid at 2.8K, but it is superficial. 
Comparing Eq.(26) and (27), one notes that $\nu (T)$ is a more appropriate 
quantity  than $\eta (T)$ for describing simply the change of the system 
around $T_{\lambda}$.  As shown in Fig.6(a), the total density  $\rho 
(T)$ of a liquid helium 4 in the normal phase 
 monotonously increases with decreasing temperature until $T=T_{\lambda}$ \cite {den}. 
  Fig.6(b) shows the kinematic shear viscosity $\nu 
(T)=\eta (T)/\rho (T)$ using $\eta (T)$  (Fig.1(a)) and $\rho (T)$ 
(Fig.6(a)),  revealing that $\nu (T)$ monotonously decreases  with 
decreasing temperature over the whole range of $T_{\lambda}<T<4.2K$.  
This means that in 1 atm,  it is just below the gas-liquid condensation 
point $T_c=4.2K$ that the strong influence of Bose statistics begins 
to suppress the shear viscosity in a liquid helium 4.

\begin{figure}
\includegraphics [scale=0.5]{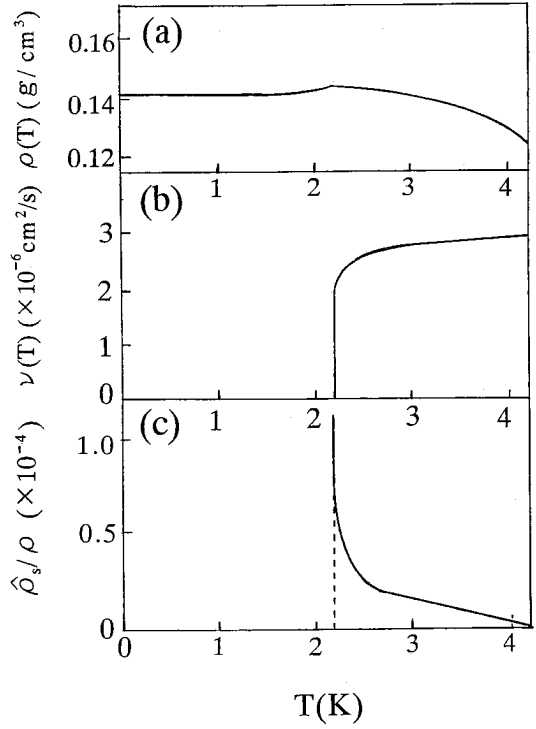}
\caption{\label{fig:epsart} (a) The temperature dependence of the total density $\rho $ of  a liquid 
  helium 4 (after the data in Ref.34). 
  (b) The temperature dependence of the kinematic shear viscosity $\nu $ of a liquid 
  helium 4 obtained using Fig.1(a) and Fig.6(a).
  (c) $\hat {\rho¥}_s(T)/\rho$ obtained by Eq.(43) using $\nu (T)$ in 
  Fig.6(b) and $\omega _s=5 \times 10^{-3} rad/s$. }
\end{figure}

This temperature  dependence and the smallness of $\nu (T)$ implies
that a liquid helium 4 in the normal phase is already an anomalous 
liquid under the strong influence of Bose statistics.  In a strict 
sense, it is not certain whether there is a completely classical regime 
in liquid helium 4 at 1 atm. But, even if we assume $T_{on}>4.2K$, it is 
impossible to analyze the onset mechanism of the decrease of $\nu $ in a 
virtual liquid at $T>4.2 K$ in 1 atm.  For the present, we assume 
$T_{on}=4.2K$, and interpret  $\nu (T)$ in Fig.6(b) using Eq.(27) with 
$\nu _n=\nu (T_{on})$.  One obtains $\hat {\rho¥}_s(T)/\rho$ by
\begin{equation}
 \frac{8}{\pi a^2\omega_s¥}¥\frac{\hat {\rho¥}_s(T)}{ \rho¥}¥ =\nu (T)^{-1}-\nu (T_{on})^{-1}.
	\label{¥}
\end{equation}¥ 
A typical value of the capillary radius used in the experiment in 
Ref.3, 4 and 5 is $a\simeq 5\times 10^{-3}cm$. 
 In the rotation experiment by Hess and Fairbank \cite {hes}, the 
 moment of inertia $I_z$ just above $T_{\lambda}$ is slightly smaller than the  normal 
 phase value $I_z^{cl}$ (see Sec.5.B).  Using these data,  Ref.14 roughly estimates
  $\hat{\rho _s}(T_{\lambda}+0.03K)/\rho\cong 8\times 10^{-5}$, and  
 $\hat{\rho _s}(T_{\lambda}+0.28K)/\rho\cong 3\times 10^{-5}$. 
 With these values and $\nu (T)$ in Fig.6(b), we obtain a rough 
 estimation of $\omega_s$ as $\omega_s \simeq 5\times 10^{-3} rad/s$ at 
 these two temperatures just above $T_{\lambda}$. 
 Figure 6(c) shows a resulting temperature dependence of $\hat {\rho¥}_s(T)/\rho$.  
 (Since the precision of these two values of $\hat {\rho¥}_s(T)/\rho$ 
 derived from currently available data is limited, the absolute value of $\hat 
 {\rho¥}_s(T)/\rho$ in Fig.6(c) has a statistical uncertainty.)

 The quantity which directly indicates the onset of superfluidity in 
 dissipative system is the change of the conductivity spectrum $\sigma (\omega)$.  In view of 
the gradual decrease of $\eta $ above $T_{\lambda}$, the sharp peak 
around $\omega =0$ and the corresponding change from $\sigma (\omega)$ to $\sigma 
_n(\omega)$ in Fig.3 must already appear at $T>T_{\lambda}$. (Since 
this change does not occur in the case of thermal fluctuations, it is 
useful for ruling out the possibility of fluctuations.) 
 To confirm this prediction, the time-resolved measurement of the oscillating flow 
velocity $\mbox{\boldmath $v$}(\mbox{\boldmath $r$},t)$ is necessary 
under the slowly oscillating pressure gradient. Such an experiment must be performed 
in a thin capillary with an inner radius of $10^{-2}\sim 10^{-1}$ mm. 
Recently,  the PIV technique recorded the velocity of tracer particles in 
a flow of liquid helium 4,  and mapped  out the velocity field  $\mbox{\boldmath 
$v$}(\mbox{\boldmath $r$})$ in a pipe \cite {don} \cite {van}. If  the PIV 
experiment is performed under the above conditions, the change of $\sigma 
(\omega)$ in Fig.3 will be observed. 
There is now ambiguity in interpretation of the movement of tracer 
particles \cite {poo}.  If tracer particles  interact only with the normal fluid flow 
and trace its velocity, the change from $\sigma (\omega)$ to $\sigma 
_n(\omega)$ as in Fig.3 will be observed.  The quantity $A$ in Eq.(10) is 
equal to the area of the shaded region $\sigma (\omega)-\sigma 
_n(\omega)$ in Fig.3(b). On the other hand, if tracer particles  interact  
with the superfluid flow as well, the emergence of the sharp peak around 
$\omega =0$ as in Fig.(b) will be detected.  In any case,  using thus 
obtained $A$, one can determine $\omega _s$  by interpreting the 
experimental data of $\eta (T)$ with Eq.(10).  Furthermore, using 
$A=2\hat {\rho¥}_s(T)/(\pi a^2)$, one will obtain 
an independent estimate of $\hat {\rho¥}_s(T)/\rho$ at $T>T_{\lambda}$.
Although such an experiment may be a difficult one, it is an experiment worth attempting.

 \section{Discussion}
\subsection{Interpretation of the $\lambda$-shape of specific heat}  
 Bose-Einstein condensation differs from ordinary phase transitions in that it 
occurs without interactions between particles.  At the early stages in 
the history of the study of phase-transition in the early part of twentieth century, almost all 
phase transitions were thought to have a $\lambda$-like-shaped 
temperature dependence of the specific heat $C(T)$ at the vicinity of $T_c$.
 As the precision of experiments was improved, however, it became clear 
that   $C(T)$ in  most phase transitions has, not the $\lambda$-like shape,
 but a $\delta$-function-like one. In a liquid helium 4, however,  
 the precise measurements  revealed that $C(T)$ intrinsically has the 
 $\lambda$-like temperature dependence. 
The principal peak at $T_{\lambda}$ has been a subject of intensive 
studies of the critical phenomena. Furthermore, the gradual rise of 
$C(T)$ above $T_{\lambda}$ has been  interpreted as a sign of enhanced 
thermal fluctuations  specific to the Bose systems near $T_{\lambda}$. 

There is a reason that such a fluctuation-based interpretation was 
accepted on the gradual rise of $C(T)$.  
In theories of the Bose system, the $V\rightarrow 
\infty$ limit is normally used. In this limit, the number density of 
particles with zero momentum is exactly zero at  $T>T_{\lambda}$, and is 
 finite only at  $T\leq T_{\lambda}$. (In the equation of states 
 $N/V=g_{3/2}(e^{\beta\mu})+V^{-1} e^{\beta\mu}/(1-e^{\beta\mu})$ 
where $g_{a}(x)=\sum_{n}x^n/n^{a}¥$, the zero momentum part 
$V^{-1} e^{\beta\mu}/(1-e^{\beta\mu})$ has a finite value at $V\rightarrow\infty$ limit only when 
 $e^{\beta\mu}/(1-e^{\beta\mu})$ diverges, that is, $\mu =0$.) 
 The $V\rightarrow \infty$ limit is necessary for the definition of the phase transition that the 
thermal average of the order parameter is exactly zero above $T_c$. 

The $V\rightarrow \infty$ limit is a good approximation only when one can 
clearly distinguish between microscopic and  macroscopic phenomena. In the Bose 
system, however,  the distinction between  ``microscopic''  and  ``macroscopic'' is 
not so obvious as in other phenomena. Since particles with zero momentum 
plays an important role  at low temperature, the coherent many-body wave function with a 
macroscopic wave length plays a significant role in microscopic descriptions of the system. 
In this sense,  ``microscopic''  and  ``macroscopic'' coexist from the 
beginning.  But, if we naively apply this $V\rightarrow \infty$ limit to the anomalous response at 
$T>T_{\lambda}$, the thermal average of the output quantity is set to be zero by $V\rightarrow 
\infty$, and we are forced to regard all of them as being caused by thermal fluctuations. 
We must carefully examine the validity of regarding a whole sample as occupying an infinite 
volume. In the thermodynamical viewpoint using  $V\rightarrow \infty$, the intermediate-sized coherent wave 
function plays a minor role above $T_{\lambda}$, but for the Bose system 
above $T_{\lambda}$, it substantially affects  the system. Hence, 
we must experimentally estimate the magnitude of effects hidden by this $V\rightarrow \infty$ limit. 
In this paper, we  viewed the gradual decrease of $\eta$ above 
$T_{\lambda}$ from  this point \cite {vie}, and showed that  
 {\it the growth of the intermediate-sized coherent wave function 
 gives rise to the decrease of $\eta $ \/}.

\subsection{Superfluidity in the dissipative and the non-dissipative systems}   
  As an example of the non-dissipative systems, we considered a rotational flow. 
In the rotational flow, the quantity directly indicating the onset of 
superfluidity is the moment of  inertia
 \begin{equation}
     I_z= I_z^{cl} \left(1-\frac{\hat {\rho _s}(T)}{\rho¥}¥\right)¥ ,¥
\end{equation}¥
where $I_z^{cl}$ is its classical value \cite {koh}. 
On the onset mechanism of superfluidity, one can see the physical 
difference between the dissipative and the non-dissipative systems in Eq.(27) and (44). 

(a) In Eq.(44), $\hat {\rho _s}(T)$ appears  as a correction to the 
coefficient of the linear term of $I_z^{cl}$. 
On the other hand, in the expansion of Eq.(27) with respect to $\nu _n$,   
$\hat {\rho _s}(T)$ appears in the non-linear higher-order terms of $\nu _n$. 
Furthermore, with decreasing temperature, the higher-order terms become dominant in Eq.(27). 

(b) In Eq.(44), the change of the transverse excitations appears in $I_z$ without being enhanced,  
 and therefore the nonzero $\hat {\rho _s}(T)$ only slightly affects $I_z$ above $T_{\lambda}$.
On the other hand, in Eq.(27), the change of the transverse excitations 
does not directly appear in $\nu $, but through the  dispersion 
integral as in Eq.(23), which is a characteristic feature of the dissipative systems. During this 
process, a sign of superfluidity in the dissipative systems is enhanced 
to an observable scale. 

(c) In Eq.(27), the peak width $\omega _s$ reflects a degree of the rigidity of superfluidity, which is 
derived from the $\omega$ dependence of $\lim_{q\to 0}[\chi^L(q,\omega)-\chi^T(q,\omega)]$, 
whereas $\omega _s$ does not appear in Eq.(44). This feature is characteristic to superfluidity in the 
dissipative systems, which manifests itself while showing the rigidity 
against the dissipation. The microscopic derivation of $\omega _s$ is a future 
problem. 

In principle, it is possible to begin with Eq.(5), and obtain Eq.(27) after perturbation 
calculations. But it may require a knowledge of the dissipation mechanism in a liquid. 
 In other words, the correct model of the dissipation mechanism of a liquid 
 must  reproduce Eq.(27) on the influence of superfluidity, because 
 Eq.(27) is based on the general argument. In this sense, one can use 
 this requirement as a criterion of the correct model of a liquid.

 As another example of superfluidity in the dissipative system, one knows 
 the anomalous thermal conductivity of a liquid helium 4 at $T<T_{\lambda}$. 
 Heat flow $q$ is expressed as $q=-\kappa\nabla T$, where $\kappa$ is the thermal conductivity 
 coefficient. At $T=T_{\lambda}$, $\kappa$ jumps to a $10^7$ times larger 
 value than $\kappa$ just above $T_{\lambda}$. At $T_{\lambda}<T<2.8K$, 
 however, the gradual rise of $\kappa$ corresponding to the gradual fall of 
 $\eta $ is not observed \cite {kel}.
 Formally, we do not know a non-dissipative phenomenon complementary to 
 the heat conduction, like the rotational flow in the shear viscosity. 
 Hence, we can not apply the formalism in 
 this paper to the onset mechanism of the anomalous thermal conductivity. 
 This difference between heat conduction and shear viscosity on a 
 formal level may have some implication on the absence of the gradual rise of 
 $\kappa$ at $T_{\lambda}<T<2.8K$.

\subsection{Comparison to  Fermi liquids}   
 In a liquid helium 3, the fall of the shear viscosity at $T_c$ is known 
as a parallel phenomenon to that of a liquid helium 4.  The formalism 
in Sec.2 is also applicable to the shear 
viscosity of a liquid helium 3.  For the behavior above $T_c$, however, 
there is a striking  difference between a  liquid helium 3 and 4. 
The phenomenon occurring in fermions at the vicinity of $T_c$ is not a 
gradual growth of the coherent wave function, but a formation of the Cooper pairs from two fermions. 
 (This difference evidently appears in the temperature dependence of the 
 specific heat: $C(T)$ of a liquid helium 3 just above $T_c$ does not 
 show the symptom of its rise to the sharp peak.)
 Once the Cooper pairs are formed, they are  composite bosons situated at low temperature 
and high density,  and immediately jumps to the ESP or BW state. 
Hence, the shear viscosity of a liquid helium 3 shows an abrupt drop at 
$T_c$ without a gradual fall above $T_c$.    
 
In electron superconductivity, the fluctuation-enhanced 
conductivity $\sigma '$  is observed above $T_c$ \cite {ttin}. In  bulk superconductors, 
however, the ratio of $\sigma '$ to the normal  conductivity  $\sigma _n$ 
is about $10^{-5}$ at the critical region, and zero outside of this region.  
 Practically,  it is unlikely that thermal fluctuations create a large 
 change of $\sigma$ at temperatures outside of the critical region.

\appendix 
\section{Maxwell's relation}
Consider the shear transformation of a solid and of a liquid. In a solid, the 
shear stress $F_{xy}$ is proportional to a shear angle $\phi$ as 
$F_{xy}=G\phi$, where $G$ is the modulus of rigidity. The value of $G$ is 
determined by dynamical processes in which vacancies in a solid move to neighboring 
positions over the energy barriers. As  $\phi$ increases, $F_{xy}$ 
increases as follows,
\begin{equation}
 \frac{dF_{xy}}{dt}=G\frac{d\phi}{dt¥}¥¥.
	\label{¥}
\end{equation}¥
In a liquid, the flow motion rearranges the relative position of particles, 
reducing the shear stress $F_{xy}$ to a certain value. 
Presumably, the rate of such a relaxation is proportional to the magnitude of $F_{xy}$, and one obtains
\begin{equation}
 \frac{dF_{xy}}{dt}=G\frac{d\phi}{dt¥}-\frac{F_{xy}}{\tau¥}¥¥¥.
	\label{¥}
\end{equation}¥
In the stationary flow after relaxation, $F_{xy}$ remains constant, and 
one obtains
\begin{equation}
 G\frac{d\phi}{dt¥}=\frac{F_{xy}}{\tau¥}¥¥¥.
	\label{¥}
\end{equation}¥
 Figure.7 shows two particles 1 and 2, each of which starts at $(x,y)$ and $(x,y+\Delta y)$ 
simultaneously and moving along the $x$-direction. Consider  a
velocity gradient $v_x(y)$ along $y$ direction. After $\Delta t$ has 
passed, they (1' and 2') are at a distance of $\Delta v_x\Delta t$ along the 
$x$-direction. As a result, the shear 
angle increases from zero to $\Delta\phi$, which satisfies $\Delta v_x\Delta 
t=\Delta y\Delta\phi$ as depicted in Fig.7. Hence, we obtain
\begin{equation}
 \frac{\partial v_x}{\partial y¥}=\frac{d\phi}{dt¥}¥.
	\label{¥}
\end{equation}¥
Using Eq.(A4) in Eq.(A3), and comparing it with Eq.(1), one obtains 
$\eta=G\tau$ (Maxwell's relation).

 \begin{figure}
\includegraphics [scale=0.5]{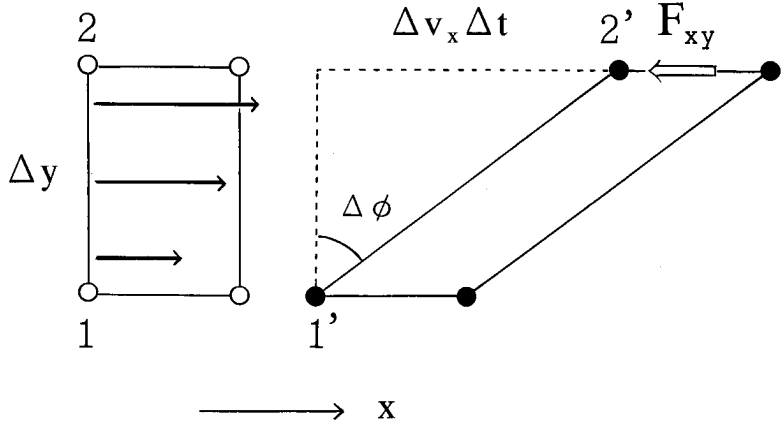}
\caption{\label{fig:epsart} In a liquid  flowing along the $x$-direction, owing to the
 velocity gradient along the $y$-direction, a small part  of a liquid with a rectangular
  shape transforms to a shape whose  cross section is a parallelogram. }
\end{figure}

 \section{Conductivity spectrum $\sigma (\omega)$}
The Navier-Stokes equation under the oscillating pressure gradient $\Delta Pe^{i\omega t}/L$ 
is written in the cylindrical polar coordinate as follows
 \begin{equation}
 \frac{\partial v}{\partial t¥}¥=\nu \left( \frac{\partial }{\partial r^2¥}+ \frac{\partial }{r\partial r¥}\right)v
                              + \frac{\Delta Pe^{i\omega t}}{\rho¥L}¥.
\end{equation}¥
The velocity field has the following form
\begin{equation}
 v(r,t)¥=\frac{\Delta Pe^{i\omega t}}{i\omega \rho¥L}+\Delta v(r,t)¥,
\end{equation}¥
with the boundary condition of $v(a,t)=0$. $\Delta v(r,t)$ satisfies 
\begin{equation}
 \frac{\partial \Delta v(r,t)}{\partial t¥}¥=\nu \left( \frac{\partial}{\partial r^2¥}
                      + \frac{\partial }{r\partial r¥}\right) \Delta v(r,t)¥,
\end{equation}¥
and therefore $\Delta v(r,t)$ has a form of Bessel function $J_0(i\lambda r)$
with $\lambda =(1+i)\sqrt {\omega /(2\nu)}$. Hence,
\begin{equation}
 v(r,t)=\frac{\Delta Pe^{i\omega t}}{i\omega \rho¥L¥}¥
                       \left(1-\frac{J_0(i\lambda r)}{J_0(i\lambda a)¥}\right)¥.
\end{equation}¥
At $r=0$, the conductivity $\sigma (\omega )$ satisfying $\rho 
v(0,t)=\sigma (\omega )a^2 \Delta Pe^{i\omega t}/L$ is given by
\begin{equation}
 \sigma (\omega )=\frac{1}{i\omega a^2¥}\left(1 
 -\frac{1}{J_0(ia(1+i)\displaystyle {\sqrt {\frac{\omega}{2\nu¥}¥}})¥}\right)¥.
\end{equation}¥
Re $\sigma (\omega)$  is schematically illustrated in Fig.3(a). (Re $\sigma (0)$ in 
Eq.(B5) agrees with Eq.(6). Im $\sigma (\omega)$ gives an expression of $\sigma 
_2(\omega )$ in Eq.(11), but it does not agree with $\sigma _2(0,\omega )$ 
in Eq.(20), because it is derived from the phenomenological 
 equation with dissipation like the Navier-Stokes equation.)

\newpage 

\begin{thebibliography}{99}
\bibitem{cur} As a response to $\mbox{\boldmath $P$}/L$, 
           $\mbox{\boldmath $j$}(\mbox{\boldmath $r$})$  at another 
           point in a flow is possible, but it results in the 
           same form of $\eta$.
\bibitem{sum} Equation (8) is the oscillator-strength sum rule 
           in terms of the conductivity.
\bibitem{vec} The particle density is adjusted to 1. In Ref.8, the definition of the vector 
         potential $\partial A/\partial t=-E$ plays the same role as Eq.(14).
\bibitem{flu}  In the fluctuation theory, the 
          correlation function of deviations from a mean value  $x-\bar {x}$ gives 
          a susceptibility around $T_c$,  an average of which is taken with the Gauss distribution.  
           Owing to the flatness of the minimum of the 
           thermodynamic  potential near $T_c$, the susceptibility has $(1-T/T_c)$ 
           dependence.  On the other hand, the susceptibility 
           (Eq.(17) and (31)) is given as the correlation function of a total quantity $x$.
            Its average is taken with the Bose-Einstein distribution. Hence, the 
           result has no  $(1-T/T_c)$ dependence.
\bibitem{wha} Whatever complicated approximation is made for 
         the susceptibility, the treatment in Eq.(19) enables us to easily derive the superfluid density from it.
\bibitem{def} Our ordinary use of $\rho$ as a longitudinal 
         susceptibility satisfying $\rho=mn=\chi^L(q,0)$, and the 
         appearance of $\rho$ in the conductivity $\sigma_{1n}=\rho 
         /(4\eta )$ as a transverse response are made compatible  
           by the fact that a replacement of $\chi^T$ by $\chi^L$ is possible in Eq.(21).
\bibitem{log} For the longitudinal displacement, the large-scale 
        inhomogeneity occurs in  the particle density, and it is therefore 
        impossible to find, in the initial configuration, a particle close 
        to that particle after displacement. See Fig.1 of Ref.15.
\bibitem{ord} In an ordinary liquid, $\tau $ is determined by the
        individual motion of particles. Hence, it is not as small as  
        $\tau $ of the collective relaxation by the coherent wave function in a superfluid.
\bibitem{com} In addition to the  bubble, it is possible that  more complex 
            diagrams, featuring intense excitations in a liquid,  exchange 
             particle lines with the tensor $J_{\mu}J_{\nu}$ in Fig.5(a). 
              While it is difficult to estimate an infinite sum of these diagrams, it adds only a small 
            correction to the onset mechanism of the decrease of $\eta $.
\bibitem{she} $\eta $ of a classical liquid is inversely proportional to the rate of
               a process in which a vacancy in a liquid propagates from one point to 
               another over the energy barriers. With decreasing temperature, this rate 
               decreases, thus increasing  $\eta $.
\bibitem{vie} The magnitude of the decrease of $\eta $ in a 
           capillary flow around $T_{\lambda}$ is strongly dependent on the radius of a capillary, 
           suggesting  that the distinction between ``microscopic'' and 
           ``macroscopic'' is obscure, and  exhibiting a feature of mesoscopic physics.
\end{thebibliography}

\end{document}